%% file: main.tex
\documentclass[aps,prl,twocolumn,preprintnumbers,amsmath,amssymb,superscriptaddress]{revtex4}
\usepackage{CJK}
\usepackage{graphicx}
\usepackage[caption=false]{subfig}
\usepackage{epsfig}
\usepackage{dcolumn}
\usepackage{bm}
\usepackage{color}
\usepackage{float}
\usepackage[colorlinks]{hyperref}
\usepackage{verbatim}
\usepackage{algorithm}
\usepackage{algorithmic}
\usepackage{tikz}
\usetikzlibrary{quantikz}

\hypersetup{citecolor = blue}

\def\bra#1{\left\langle#1\right|}
\def\ket#1{\left|#1\right\rangle}

\def\be{\begin{equation}}       \def\ee{\end{equation}}
\def\bea{\begin{eqnarray}}      \def\eea{\end{eqnarray}}
\def\ba{\begin{array}}
\def\ea{\end{array}}
\def\bnum{\begin{enumerate} }
\def\enum{\end{enumerate}}

\def\=>{\Rightarrow}
\def\>{\rightarrow}

\def\eye2{Fathbb{I}}

\renewcommand{\>}{\rangle}

\renewcommand{\rm}[1]{\mathrm{#1}}

\newcommand{\ipic}[4]{
	\begin{figure}[!th]\centering
		\includegraphics[#4]{#1}
		\caption{#2}  \label{#3}
	\end{figure}
}

\usepackage{listings}
\usepackage{color}
\definecolor{lightgray}{gray}{1}

\begin{document}

\title{Neural network encoded variational quantum algorithms}

\author{Jiaqi Miao}
\affiliation{School of Physics, Zhejiang University, Hangzhou, Zhejiang 310000, China}

\author{Chang-Yu Hsieh}
\email{kimhsieh@zju.edu.cn}
\affiliation{Innovation Institute for Artificial Intelligence in Medicine of Zhejiang University, College of Pharmaceutical Sciences, Zhejiang University, Hangzhou, 310058, China}

\author{Shi-Xin Zhang}
\email{shixinzhang@tencent.com}
\affiliation{Tencent Quantum Laboratory, Tencent, Shenzhen, Guangdong 518057, China}

\date{\today}

\begin{abstract}

    We introduce a general framework called neural network (NN) encoded variational quantum algorithms (VQAs), or NN-VQA for short, to address the challenges of implementing VQAs on noisy intermediate-scale quantum (NISQ) computers. Specifically, NN-VQA feeds input (such as parameters of a Hamiltonian) from a given problem to a neural network and uses its outputs to parameterize an ansatz circuit for the standard VQA. Combining the strengths of NN and parameterized quantum circuits, NN-VQA can dramatically accelerate the training process of VQAs and handle a broad family of related problems with varying input parameters with the pre-trained NN. To concretely illustrate the merits of NN-VQA, we present results on NN-variational quantum eigensolver (VQE) for solving the ground state of parameterized XXZ spin models. 
    Our results demonstrate that NN-VQE is able to estimate the ground-state energies of parameterized Hamiltonians with high precision without fine-tuning, and significantly reduce the overall training cost to estimate ground-state properties across the phases of XXZ Hamiltonian. We also employ an active-learning strategy to further increase the training efficiency while maintaining prediction accuracy. 
    These encouraging results demonstrate that NN-VQAs offer a new hybrid quantum-classical paradigm to utilize NISQ resources for solving more realistic and challenging computational problems. 

\end{abstract}

\maketitle

    \textit{Introduction.} -- 
    Today's noisy intermediate-scale quantum (NISQ) computers \cite{Preskill2018} are far from delivering an unambiguous quantum advantage. Variational quantum algorithms (VQAs), as one of the most representative algorithm primitives in the NISQ era \cite{Bharti2022, Cerezo2020b, Endo2020, Tilly2021}, utilize a quantum-classical hybrid scheme, where the quantum processor prepares target quantum states and measurement is made to extract useful information for the classical computer to explore and optimize. VQAs have now been widely applied to solve quantum optimization, quantum simulation, and quantum machine learning problems \cite{Peruzzo2014, McClean2016, farhi2014, KAK1995, pylkkanen1995, Kerstin2020, Cong2019, Maria2020, Romero2017, Benedetti2019, Cheng2023a}.
   
    Among various VQAs, the variational quantum eigensolver (VQE) \cite{Peruzzo2014, McClean2016} certainly stands out as one of the most exemplary algorithms. VQE employs Rayleigh-Ritz variational principle to approximate the ground state of a given Hamiltonian $\hat{H}$ with a parameterized quantum circuit (PQC). Many studies on the strengths and fundamental limitations of VQAs are first systematically investigated and revealed by studying how VQE performs in different contexts. Despite some early hopes of VQAs' potential quantum advantages in addressing some realistic computational problems, this goal still remains elusive. In fact, it is now known that the current formulation of the vanilla VQAs faces way too many obstacles 
    for them to deliver any practical advantages.
    
    
    There is a pressing need to develop novel hybrid quantum-classical approachs to better utilize the full power of quantum computational resources while avoiding as many shortcomings of the vanilla VQAs as possible. For instance, a core problem of the standard VQA is to identify the suitable circuit parameters for a given problem, i.e. the optimization or training procedure. From a practical perspective, the training procedure often takes many steps which leads to a large budget for measurement shots. Besides, the training procedure could be more sensitive to noise and decoherence compared to the inference procedure. Therefore, training of VQAs is expensive as it must be conducted on very high-quality quantum devices with a large budget of measurement shots.
    
    In terms of theoretical perspective, the difficulties associated with the optimization of VQAs stem from at least two fundamental obstacles.
    One severe challenge is the phenomenon of vanishing gradients named barren plateaus (BPs) \cite{McClean2018, Cerezo2021BP, Cerezo2021HOBP, Wang2021, Arrasmith2022}. 
    Though there are many attempts to mitigate BP issues \cite{kashif2023, miao2023, park2023, robertson2022, Mele2022, liu2022, Friedrich2022, kulshrestha2022, broers2022, Pesah2021, Liu2023a}, the occurrence of BP, in general, implies that exponential quantum resources are required to navigate through the exponentially flattened cost function landscape $C(\theta)$, which could negate the potential quantum advantages of VQAs. 
    Another related problem for VQAs' non-convexity energy landscape is the occurrence of many local minima \cite{Bittel2021, Anschuetz2022}, which can easily trap the training trajectories.

\ipic{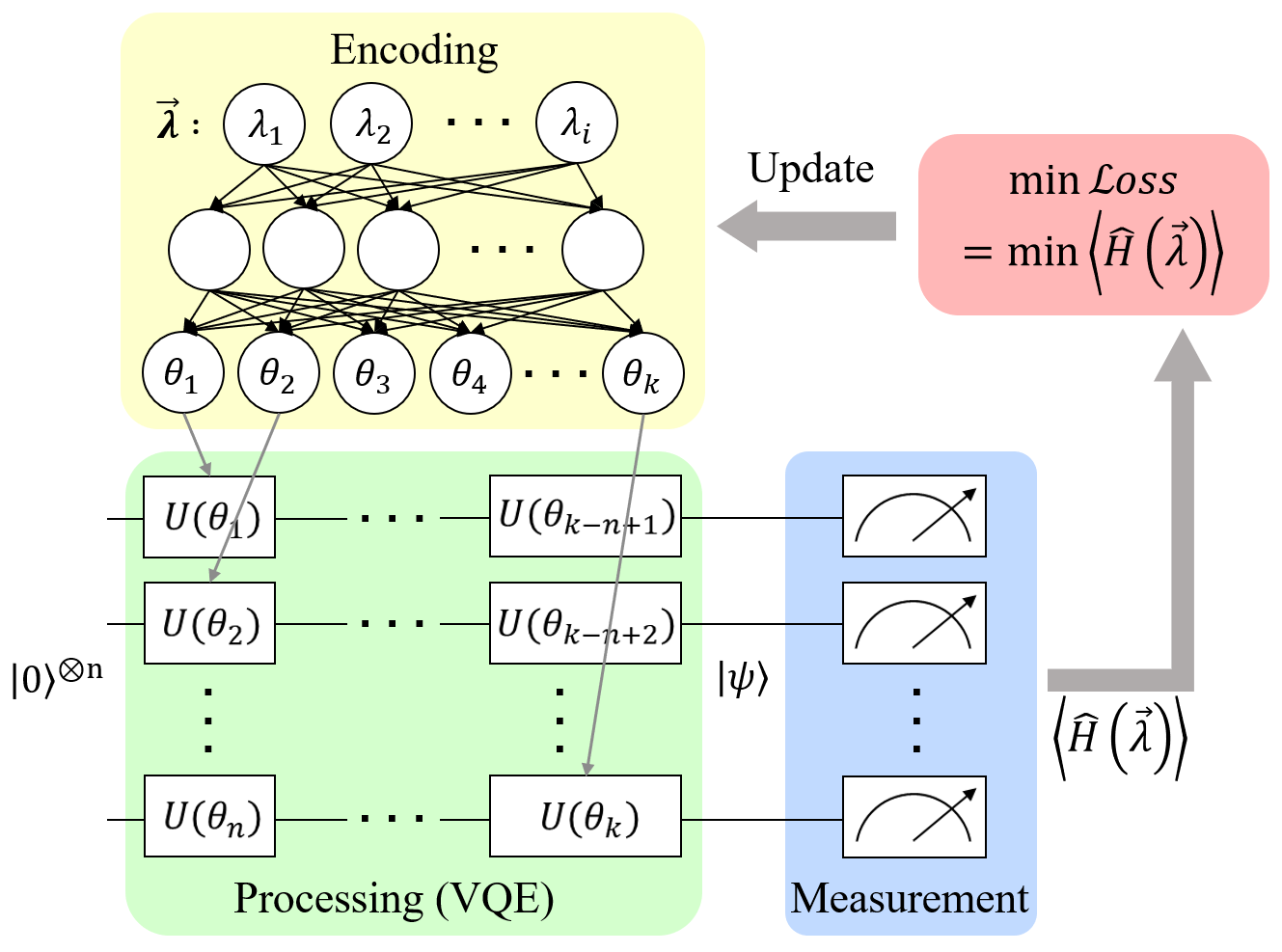}{Schematic workflow for NN-VQE. The Hamiltonian parameters $\bm{\lambda}$ are the input of the encoder neural network, which produces the parameterized quantum circuit (PQC) parameters $\bm{\theta}$ as output. The PQC, parameterized by $\bm{\theta}$, is then used in the processing module of the VQE to prepare an output state $\ket{\psi} = U(\bm{\theta}) \ket{\bm{0}}$, where $\ket{\bm{0}}$ is the initial state. The cost function can be estimated according to Eq. (\ref{eq3}), and the weights in the neural network are optimized using a gradient-based optimizer.}{fig:workflow}{width=0.95\linewidth}

    In the plain VQA setups, application problems are optimized and solved instance by instance with the same circuit structure, namely, we need to retrain the model for each instance. This workflow renders the optimization issues discussed above more detrimental in the VQA context. 
    Therefore, a general framework to solve the parameterized problem instances jointly and to separate the pre-training process from the inference process is highly desired. Such a framework would address the optimization bottlenecks from two angles. For the pre-training procedure, the joint training on multiple problem instances speeds up the optimization convergence by alleviating the BP and local minima issues. And for the inference procedure conducted by the end-users, there is no need to retrain or fine-tune the model so that the end-users with limited quantum resources are free from the thorny training issues.

\ipic{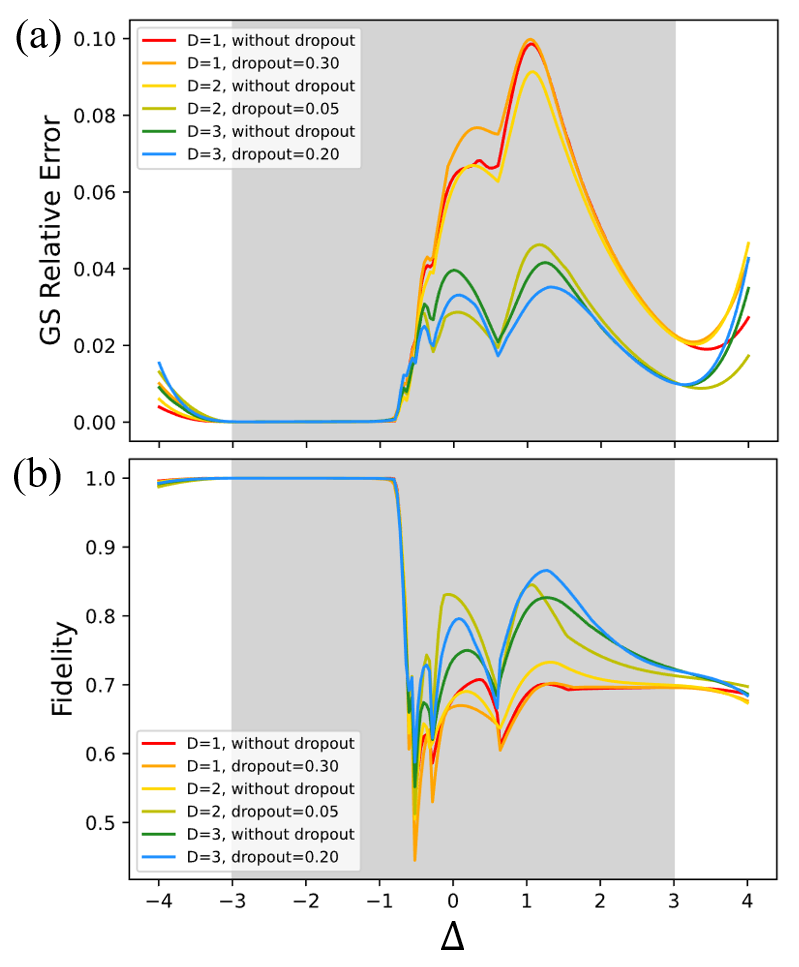}{Results on $n=8$ qubit one-tunable-parameter 1D XXZ spin chain with a transverse field strength $\lambda=0.75$ within NN-VQE framework: (a)  Relative errors of ground-state energies of different circuit block depth $D$ with and without dropout from the pre-trained model. (b) Fidelity between the output state of NN-VQE and the exact ground state.}{fig:1xxz}{width=0.8\linewidth}

    In this Letter, we introduce a general framework -- neural network encoded variational quantum algorithms (NN-VQAs). There are many works that integrate the neural network with the quantum circuit from different angles such as quantum state tomography, quantum error mitigation, quantum architecture search, and expressive capacity enhancement \cite{Torlai2018a, verdon2019, liu2019, hsieh2021, Zhang2020b, Zhang2021, Benedetti2021, Zhang2022, Friedrich2022, Zhang2021d, Bennewitz2022, Huembeli2022, deschoulepnikoff2023}. Our framework further expands the possibility of such an interplay from a new perspective. NN-VQAs successfully address all the aforementioned challenges: (i) NN-VQAs use the Hamiltonian parameters as the input to a neural network, which enables to solve a parameterized model through only a single pre-training process; (ii) the pre-trained NN-VQAs can give a good estimation with test Hamiltonians beyond the training set with good generalization capability; (iii) active learning method can be adopted to further reduce the number of training samples and thus the number of total measurement shots; (iv) NN-VQAs could significantly speedup the optimization convergence of VQAs, alleviating the issues of BP and local minima. Therefore, by using a neural network as the encoding module, our approach provides a good ground state approximation using only a small number of training points and greatly saves the required quantum resources. Moreover, our framework can enable the separation of training and inference and sketch a potential future interface to utilize VQAs for end-users.


    \textit{Theoretical Framework.} -- In this section, we introduce the framework of NN-VQE for ground state problems, and the framework can be similarly generalized to VQE for excited states \cite{Zhang2022ba, Santagati2018a, Liu2021ca} or other VQA scenarios.
    
    The schematic workflow for NN-VQE is shown in Fig. \ref{fig:workflow}. Given a parameterized Hamiltonian $\hat{H} = \hat{H}(\bm{\lambda})$, where $\bm{\lambda} $ consists of $p$ different Hamiltonian parameters, our aim is to solve the ground state of the parameterized Hamiltonian. We choose a subset of $\bm{\lambda}$ as the training set $\bm{\tilde{\lambda}}=\{\tilde{\bm{\lambda}}_i\}$.
   
    To train an NN-VQE, we use $\bm{\tilde{\lambda}}$ as the input of the encoding neural network, and get the output 
    \begin{equation} \label{eq1}
    \bm{\theta}_i = f_{\bm{\phi}} (\bm{\tilde{\lambda}}_i),     
    \end{equation}
    where we denote the neural network as a general parameterized function $f_{\bm{\phi}}$ with the training neural weights as $\bm{\phi}$. The number of the output of the neural network is the same as the number of the PQC parameters, and we load each neural network output element to the corresponding circuit parameters. 
    The PQC $U(\bm{\theta})$ for VQE is initialized in the $\ket{\bm{0}} = \ket{0} ^ {\otimes n}$ state. Therefore, the output target state for Hamiltonian $\hat{H}(\bm{\tilde{\lambda}}_i)$ should be
    \begin{equation} \label{eq2}
    \ket{\psi_i} = U(\bm{\theta}) \ket{\bm{0}} = U \left( f_{\bm{\phi}} (\bm{\tilde{\lambda}}_i) \right) \ket{\bm{0}}.
    \end{equation}
    The cost function for ground state VQE is the expectation of $\hat{H}(\bm{\lambda})$:
    \begin{eqnarray} \label{eq3}
    & C\left(\bm{\phi} \right)  
    = \sum_i \langle \hat{H}(\bm{\tilde{\lambda}}_i) \rangle   \nonumber \\
    & = \sum_i \bra{\bm{0}} U^\dagger \left( f_{\bm{\phi}} (\tilde{\lambda}_i) \right) \hat{H} ( \tilde{\lambda}_i ) U \left( f_{\bm{\phi}} (\tilde{\lambda}_i) \right) \ket{\bm{0}}. 
    \end{eqnarray}
    Finally, we compute the gradients with respect to the neural network (back-propagation via the PQC parameters) and minimize the cost function $C \left( \bm{\phi} \right) $ using gradient descent, obtaining the optimal weights $\bm{\phi^*}$ for the neural network. Since such a training procedure only happen once and the trained model can be used to approximate the ground state of the family of Hamiltonians, we call this stage pre-training. Upon completion of pre-training, the efficacy of the NN-VQE can be evaluated using a test set of different $\bm{\lambda}$ from the training set. 

    \textit{Results.} -- In this section, we demonstrate the effectiveness of our framework using numerical simulation with TensorCircuit \cite{Zhang2023}. The testbed model is the one-dimensional (1D) antiferromagnetic XXZ spin Hamiltonian with an external magnetic field subject to the periodic boundary conditions
    \begin{equation} \label{eq4}
        \hat{H} = \sum_{i, i+1} \left( X_i X_{i+1} + Y_i Y_{i+1} + \Delta Z_i Z_{i+1} \right) + \lambda \sum_{i} Z_i ,
    \end{equation}
    where $\Delta$ is the anisotropy parameter and $\lambda$ is the transverse field strength.

    We start from the one-parameter XXZ model with the transverse field strength fixed to $\lambda=0.75$. The training set of $\Delta$ is composed of 20 equispaced points in the interval of $[-3.0, 3.0]$. The performance of the NN-VQE is evaluated on an expanded test set consisting of 201 equispaced values of $\Delta$ in the interval of $[-4.0, 4.0]$. The circuit ansatz we use in this section is inspired by MERA \cite{Vidal2008, Evenbly2009}. Specifically, we employ deep multi-scale entanglement renormalization ansatz (DMERA) circuits \cite{Kim2017a, Sewell2022DMERA}, where $D$ is the circuit depth in each block 
    (see the SM for details). The neural network we use is a simple fully connected neural network with a dropout layer. 
    The size of the input layer is 1 corresponding to the number of Hamiltonian parameters $\Delta$, and the size of the output layer corresponds to the number of PQC parameters 
    (see the SM for the detailed neural structure).

\ipic{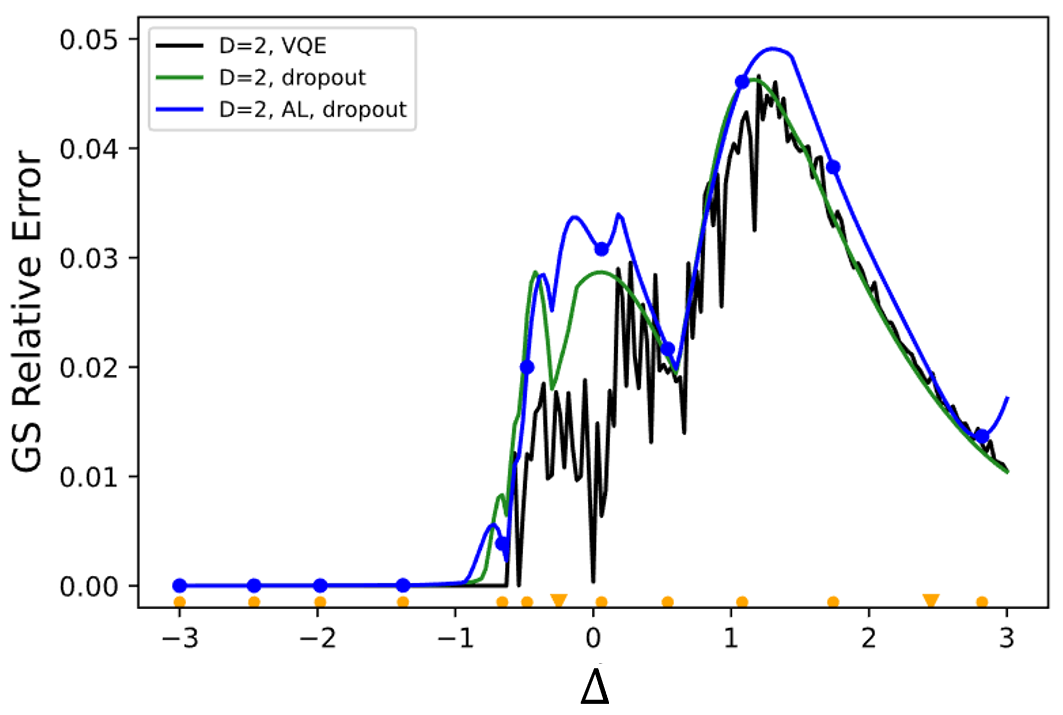}{Active learning for NN-VQE. We use MERA circuit with $D=2$. The black line is the result of VQE separately trained on each point. The green line is the NN-VQE with dropout and has the same circuit structure as the black line. The blue line shares the same structure (NN encoder and circuit ansatz) as the green line but uses active learning to reduce sample size. The training set of the green line consists of a set of equispaced 20 $\Delta$s in the interval of $[-3.0, 3.0]$ used in the previous analysis. However, by employing active learning, we use only 11 actively selected points to attain the blue line. The training set used for active learning is indicated by dots along the blue line. Remarkably, despite the reduced training set size, the blue line still exhibits a reliable estimation of the ground-state energy. The actively chosen dots are projected onto the x-axis as the orange dots, and the orange inverted triangles are the phase transition points. 
}{fig:AL}{width=0.8\linewidth}

    For the 1D XXZ spin chain consisting of 8 qubits, we pre-train the model within the NN-VQA framework and evaluate the performance with different circuit block depths $D$. The results for ground state (GS) energy prediction are shown in Fig. \ref{fig:1xxz}(a). The simulation accuracy improves with larger $D$ and dropout in the neural network. We also display the corresponding fidelities with the exact ground state in Fig. \ref{fig:1xxz}(b).  
    The results underscore the ability of the NN-VQE to effectively prepare the ground state as a function of the Hamiltonian parameters without fine-tuning or retraining.
    We note that NN-VQE demonstrates a favorable generalization capability. As shown in Fig. \ref{fig:1xxz}, in regions devoid of shadows on either side (regions of no training points), the NN-VQE still provides highly reliable estimations to some extent. 

    Compared with previous work on meta-VQE \cite{Cervera-Lierta2021}, when the PQC structures are the same, NN-VQE uses fewer quantum resources while yielding better ground-state energy estimation results (see the SM for details). Such advantages are mainly brought by the expressive power of general neural networks. 
    
    In the previous analysis, the training set is selected in an equispaced manner. Such a strategy can be improved by utilizing active learning techniques \cite{Valerii1972}. We can maintain the same level of ground-state energy accuracy while using a smaller number of training points. 

\ipic{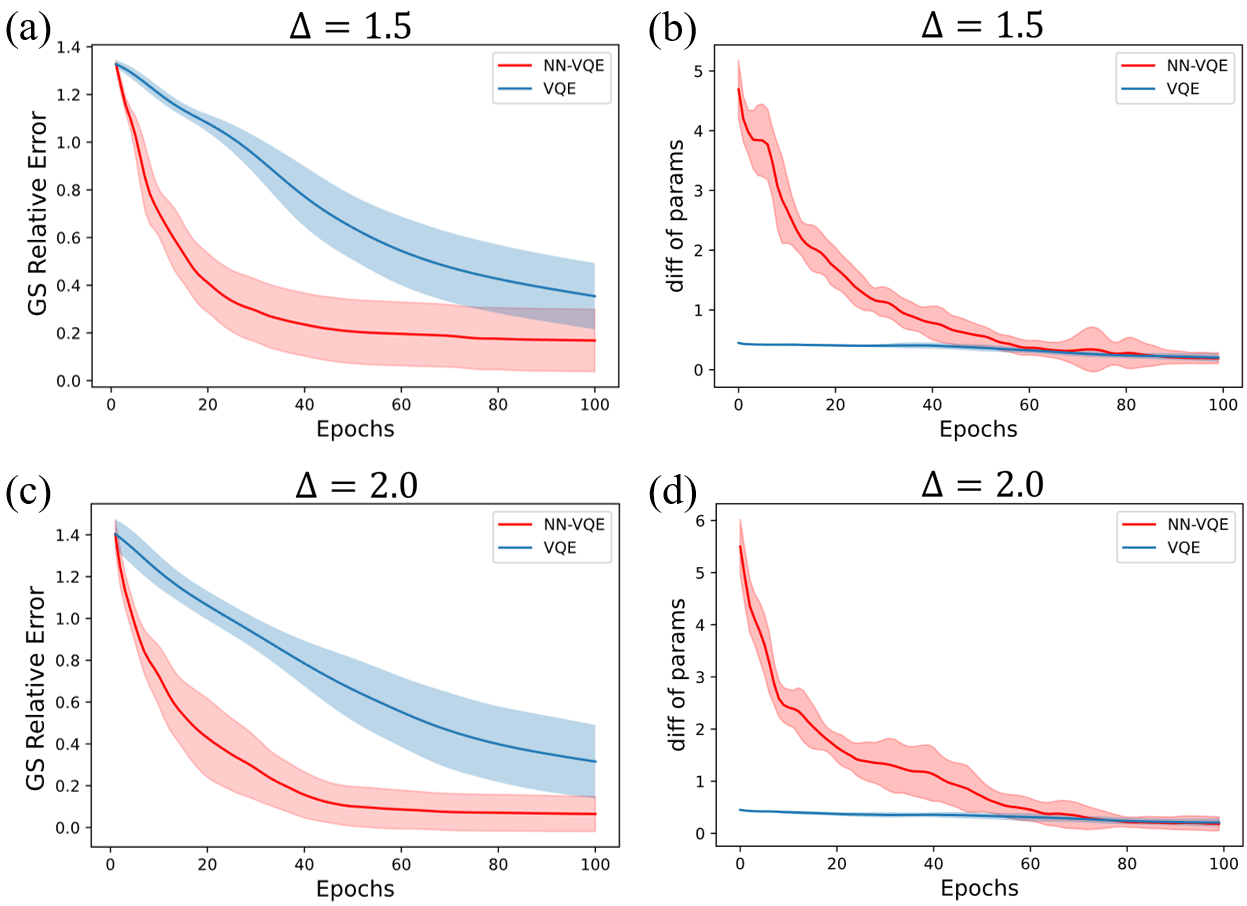}{Speedup in the optimization process of NN-VQE and the corresponding PQC parameter changes. (a)(c) The ground-state energy relative errors for an $n = 12$ XXZ spin chain when $\Delta = 1.5, 2.0$ are shown with respect to epochs. Hardware-efficient ansatz with $D = 3$ is used. The red and blue lines correspond to the ground-state energy relative errors and standard deviation of NN-VQE and VQE respectively. (b)(d) The parameter differences when training VQE and NN-VQE in corresponding $\Delta$. The difference is the sum of the absolute value of parameter differences between epochs. NN-VQE brings a more dramatic circuit parameter change at the beginning of the optimization process, which speedups the optimization process.}{fig:speedup}{width=\linewidth}

    Various active learning schemes can be easily incorporated into the NN-VQE. For example, we begin by randomly selecting one point from the $\Delta$ pool $P$ as the initial training set. We train the NN-VQE based on the training set and get the neural network weights ${\bm{\phi^*}}$. Obviously, the training set $P^*$ is a subset of the pool $P$. Subsequently, we calculate the acquisition function specifically designed in this scenario. The active learning acquisition function in our problem is defined as
    \begin{equation} \label{eq5}
    C_{AL} 
    = \langle \hat{H}^2 (\Delta) \rangle _ {\bm{\phi^*}} -  \langle \hat{H} (\Delta) \rangle _ {\bm{\phi^*}}  ^2 + \mu \min{| \Delta - \bm{\Delta^*} |} ,
    \end{equation}
    where $\Delta \in P$, $\Delta^* \in P^*$, $\mu$ a preset hyperparameter. The first two terms are the variance of the Hamiltonian $\hat{H}(\Delta)$ with ${\bm{\phi^*}}$ trained on the training set. In the last term, we first calculate the distance between $\Delta$ and all $\bm{\Delta^*}$ in the training set and find the minimum distance. We employ the hyperparameter $\mu$ to find a large variance but prevent a close point from being chosen. The two terms reflect the exploitation and exploration trade-off of the active learning technique. We add the $\Delta$ with the largest $C_{AL}(\Delta)$ to the training set. Iteratively, we repeat this process of expanding the training set until the test relative error of the ground-state energy falls below a predetermined threshold. 


    By this method, we obtain a training set consisting of 11 points. The corresponding results are shown in Fig. \ref{fig:AL}. Remarkably, even with a training set size that is only half of the previous set, the model still gives a reliable estimation of the ground-state energy. Moreover, when we visualize the training sets (see the orange dots in Fig. \ref{fig:AL}), we find them nearly equispaced except for the points near the phase transition point of the Hamiltonian (see the orange inverted triangles in Fig. \ref{fig:AL}). This observation roughly corresponds to an intuition that the ground-state wavefunction might experience a more dramatic change around the phase transition point which requires more training points to better capture.

\ipic{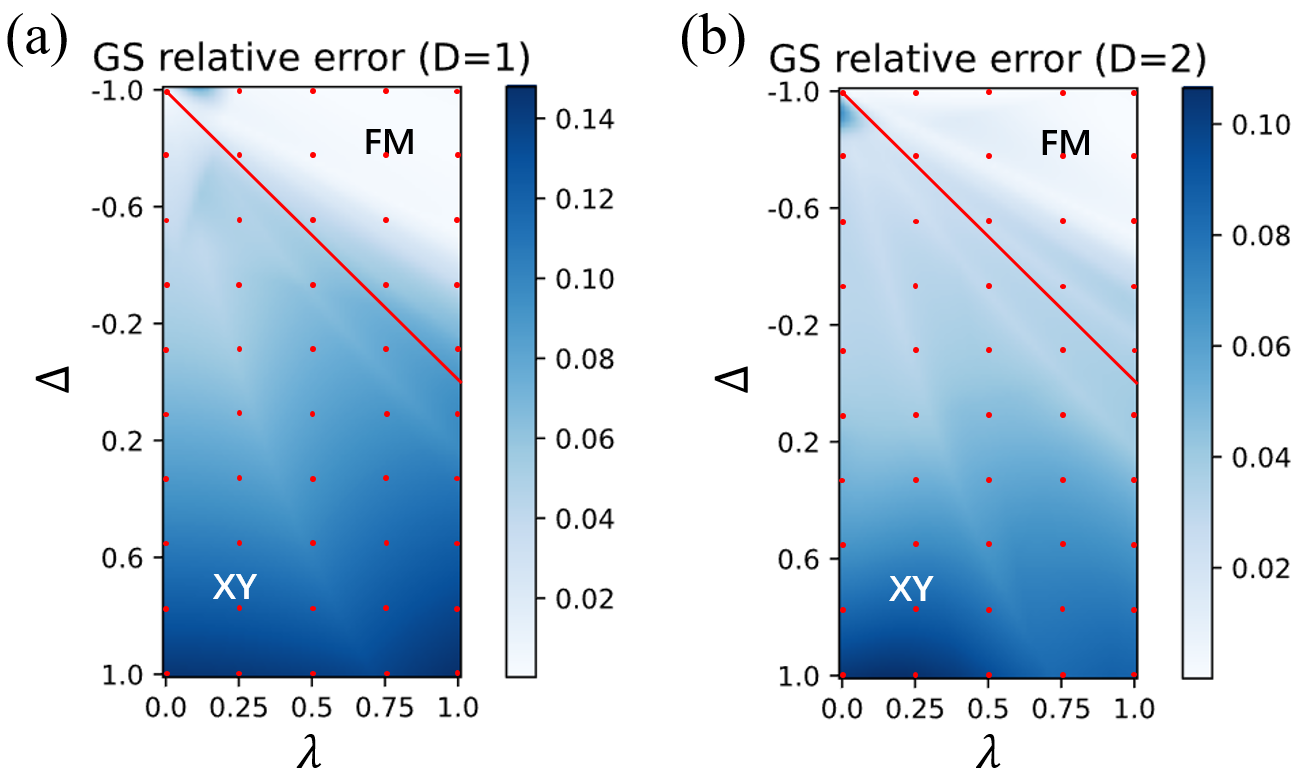}{Relative errors of ground-state energies for an $n=12$ 1D XXZ spin chain with two tunable Hamiltonian parameters, using hardware-efficient ansatz with circuit depth $D = 1, 2$. The red dots are the training set. The red lines are the exact phase transition line between the ferromagnetic phase (FM) and the XY phase \cite{braiorrorrs2015}.}{fig:2xxz}{width=\linewidth}

    Another remarkable advantage of NN-VQE is the training efficiency. As shown in Fig. \ref{fig:speedup} (a)(c), NN-VQE has a significant speedup in the optimization procedure compared with plain VQE. The energy cost function drops more rapidly, which offers great benefits for NISQ computers since fewer training epochs and thus fewer quantum resources are required. 
    Such advantages benefit from the NN-PQC hybrid architecture. The neural network brings a more dramatic change in the PQC parameters at the beginning stage of the optimization process as shown in Fig. \ref{fig:speedup} (b)(d), which might also be relevant in mitigating the BP issue.

    In order to demonstrate the effectiveness of the NN-VQE in estimating a multiparameter Hamiltonian, we extend our study to the two-parameter XXZ model. In this model, both the anisotropy parameter $\Delta$ and the transverse field strength $\lambda$ are tunable in the Hamiltonian in Eq. \ref{eq4}. The training set for $\Delta$ consists of 10 equispaced points in the interval of $[-1.0, 1.0]$, while for $\lambda$ consists of 5 equispaced points in the interval of $[0.0, 1.0]$. 
    The ansatz used is the hardware-efficient ansatz \cite{Kandala2017} with two-qubit gates in the ladder layout of depth $D$ (see the SM for details). 
    The encoding neural network also shares a similar structure as the one-parameter case but now the input takes two values $\Delta$ and $\lambda$. 

    The numerical results are presented in Fig. \ref{fig:2xxz}. Remarkably, the NN-VQE, using a neural network with two inputs, yields excellent performance in estimating the ground state across different phases. 
    This result implies the robustness and versatility of the NN-VQE in simulating complex quantum systems governed by a multiple-parameter Hamiltonian.


    \textit{Discussion.} -- In this Letter, we introduce the NN-VQA framework. More specifically, we first use a neural network to transform the Hamiltonian parameters to the optimized parameters in the PQC for VQA. We show the validity and effectiveness of the framework in solving the XXZ Hamiltonian ground state with different parameters through only one pre-training procedure without any problem instance specific fine-tuning. In order to further reduce the pre-training overhead, we also employ an active learning heuristic where the progressively built training set can be greatly reduced. We also find that the NN-VQE pipeline can speed up the training process.


    In terms of the neural network part, we can introduce more physics-inspired neural network structures for multi-parameter Hamiltonian VQE problems. For example, considering the random Ising model where the couplings at each bond or site are different, we can abstract the Hamiltonian parameters as a graph where the node and edge weights describe the Hamiltonian form. In such cases, we believe a graph neural network (GNN) \cite{Franco2009, Alessio2009} is more suitable for the encoding task as the symmetry and geometry can also be properly addressed in a well-designed GNN. And the power of considering local geometry as in the GNN approach is proven to be exponentially sample efficient in learning quantum state properties \cite{Huang2021b, onorati2023, lewis2023, che2023}. 

    Our framework envisions a future paradigm to utilize quantum computers. The encoding neural network can be pre-trained on high-quality quantum devices with a large time and measurement budget. The pre-trained model can be efficiently saved on classical computers and shared via the cloud. Since the NN-VQE can be targeted to a large family of quantum systems that can be connected via lots of parameters, a large pre-trained model could be of general interest for solving various problems. The end-users can download the large pre-trained classical model and extract the trained circuit parameters given the specific problem they are interested in solving. In this paradigm, the end users are free from training on quantum computers and can utilize the power of quantum computers more efficiently. It is also worth noting that at the training stage, due to the nature of multiple training points, it is very easy to utilize the data parallelism and pre-train the NN-VQE with many quantum computers.



\emph{Acknowledgements: }
    We gratefully thank Gaoxiang Ye for useful discussions.

\input{main.bbl}
\clearpage

\begin{widetext}
    \section*{Supplemental Materials}
        \renewcommand{\theequation}{S\arabic{equation}}
        \setcounter{equation}{0}
        \renewcommand{\thefigure}{S\arabic{figure}}
        \setcounter{figure}{0}
        \renewcommand{\thetable}{S\arabic{table}}
        \setcounter{table}{0}

        \subsection{Notations for the model, the variational circuit, and the neural network}

            \textit{The Hamiltonian}. The Hamiltonian used is a 1D XXZ model with periodic boundary conditions, with the transverse field strength $\lambda$ and the anisotropy parameter $\Delta$:
            \begin{equation}
                \hat{H} = \sum_{i, i+1} \left( X_i X_{i+1} + Y_i Y_{i+1} + \Delta Z_i Z_{i+1} \right) + \lambda \sum_{i} Z_i .
            \end{equation}
            When we use the single-parameter 1D XXZ model, we fix the transverse field strength to $\lambda = 0.75$, and the anisotropy parameter $\Delta$ is chosen as the varying parameter. When we use the two-parameter 1D XXZ model, both $\lambda$ and $\Delta$ are varying parameters.

            \textit{The circuit ansatz.} There are two main circuit ansatzes employed in this work: the ladder-wise hardware efficient ansatz (HEA) and the MERA ansatz.

            The hardware-efficient ansatz \cite{Kandala2017} encompasses a range of ansatzes that are directly tailored to the given quantum hardware employed in the experiment, avoiding the circuit depth overhead arising from transforming an arbitrary unitary into a sequence of local native gates. We employ ladder-wise HEA (see Fig. \ref{fig:supp_hea_circuit}) in this Letter. The representation of the ladder-wise HEA can be expressed as
            \begin{equation}
                \ket{\psi_D} = U \ket{\psi_{D-1}},
            \end{equation}            
            where $U = \prod_{i,j} R_2 (\bm{\theta_{ij}}) \times \prod_{i} R_1 (\bm{\theta_{i}})$ is a block of unitary (see Fig. \ref{fig:supp_hea_circuit} as a block of ladder-wise HEA unitary), and $D$ is the circuit depth representing the number of repetitions of the block. In each block, parameterized Rx and Rz gates are used as single-qubit gates $R_1(\bm{\theta_i}) = R_z(\theta_{i1}) R_x(\theta_{i2})$, and parameterized Rxx, Ryy, and Rzz gates are used as two-qubit gates $R_2(\bm{\theta_{ij}}) = R_{yy}(\theta_{ij1}) R_{xx}(\theta_{ij2}) R_{zz}(\theta_{ij3})$ arranged in a ladder-wise pattern. Before all the blocks, single-qubit rotation gates $R'_1(\bm{\theta_i}) = R_x(\theta_{i1}) R_z(\theta_{i2}) R_x(\theta_{i3})$ are operated on all the qubits, transferring the initial state $\ket{\bm{0}}$ to $\ket{\psi_0} = R'_1(\bm{\theta_i})\ket{\bm{0}}$.
            
            The multi-scale entanglement renormalization ansatz (MERA) tensor network (see Fig. \ref{fig:supp_mera_circuit}) can be adopted as a viable circuit ansatz for VQE simulations. MERA starts from a single qubit in the $\ket{0}$ state and progressively enlarges the Hilbert space by tensoring additional qubits in the $\ket{0}$ state \cite{Vidal2008, Evenbly2009}. The tree-like MERA tensor network constructed by this progressive enlargement corresponds to successively coarse-grained states and ultimately implements a fine-grained scaling transformation. The scale transformation is
            \begin{equation}
                \ket{\psi_{l+1}} = U_l (\theta) \left( \ket{\psi_{l}} \otimes \ket{0}^{\otimes2^l} \right),
            \end{equation}
            where $2^l$ is the number of fresh qubits introduced in the $l$-th layer and $U_{l} = \prod_{i=2}^{2l} R_2 (\bm{\theta_{i,i+1}}) \times \prod_{j=1}^{2l-1} R_2 (\bm{\theta_{j,j+1}}) \times \prod_{k=1}^{2l} R_1 (\bm{\theta_{k}})$ (all $i$ are even and all $j$ are odd). We denote $D$ as the depth of the brickwork unitaries $U_l$. For example, we set $D=2$ in Fig. \ref{fig:supp_mera_circuit}. There are $D=2$ blocks (yellow(1) and green(2)) in each $U_l$, and there are 2 brickwork unitaries $U_1$ and $U_2$ expanding the MERA network to 4 qubits. In each block, parameterized Rx and Rz gates are used as single-qubit gates $R_1(\bm{\theta_i}) = R_z(\theta_{i1}) R_x(\theta_{i2})$, and parameterized Rxx and Rzz gates are used as two-qubit gates $R_2(\bm{\theta_{ij}}) = R_{zz}(\theta_{ij1}) R_{xx}(\theta_{ij2})$. And before all the brickwork unitaries, single-qubit rotation gates $R'_1(\bm{\theta_i}) = R_x(\theta_{i1}) R_z(\theta_{i2}) R_x(\theta_{i3})$ are operated on all the qubits.

            \textit{The encoding neural network.} The neural network of the NN-VQE has one input layer, one hidden layer, and one output layer, which are all fully connected linear layers. The node number of the input layer corresponds to the number of Hamiltonian parameters, the node number of the hidden layer is a hyperparameter that varies with the circuit depth $D$, and the node number of the output layer corresponds to the number of PQC parameters. We also employ dropout layers after the hidden layer to avoid overfitting. To initialize the neural network, we choose the normal random initialization with the mean value $= 0.0$ and the standard deviation $= 0.1$. The optimizer we used in gradient descent is Adam and the learning schedule is the hyperparameter tuned for each case.

\ipic{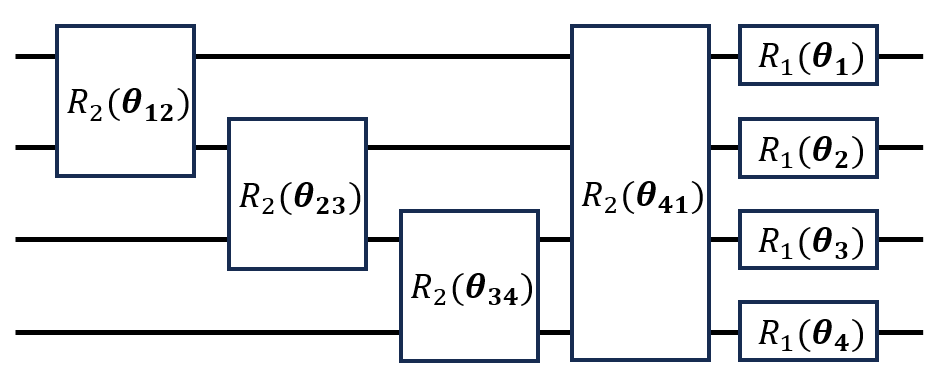}{A block of ladder-wise hardware efficient ansatz circuit with 4 qubits ($D = 1$). Each one-qubit gate $R_1(\bm{\theta_i})$ corresponds to $R_z(\theta_{i1}) R_x(\theta_{i2})$ and two-qubit gate $R_2(\bm{\theta_{ij}})$ corresponds to $R_{zz}(\theta_{ij1}) R_{xx}(\theta_{ij2}) R_{yy}(\theta_{ij3})$. }{fig:supp_hea_circuit}{width=0.53\linewidth}

\ipic{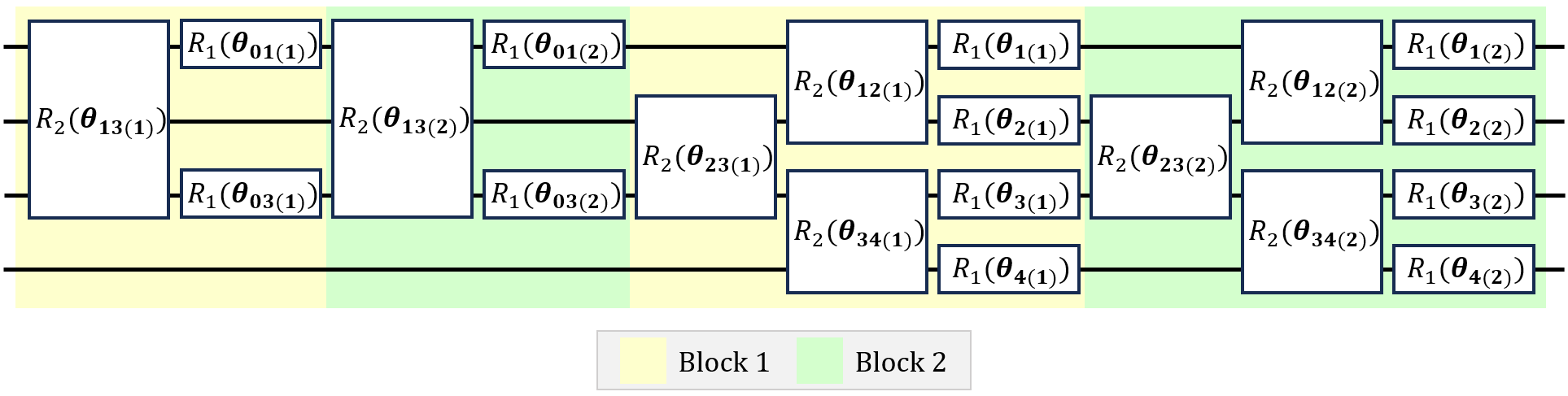}{Two blocks of MERA circuit with 4 qubits ($D = 2$). The yellow-shaded circuit is block 1 and the green-shaded circuit is block 2. In each block, there is a complete MERA structure. Each one-qubit gate $R_1(\bm{\theta_i})$ corresponds to $R_z(\theta_{i1}) R_x(\theta_{i2})$ and the two-qubit gate $R_2(\bm{\theta_{ij}})$ corresponds to $R_{zz}(\theta_{ij1}) R_{xx}(\theta_{ij2})$. All parameters are independently optimized with no parameter sharing.}{fig:supp_mera_circuit}{width=\linewidth}

\ipic{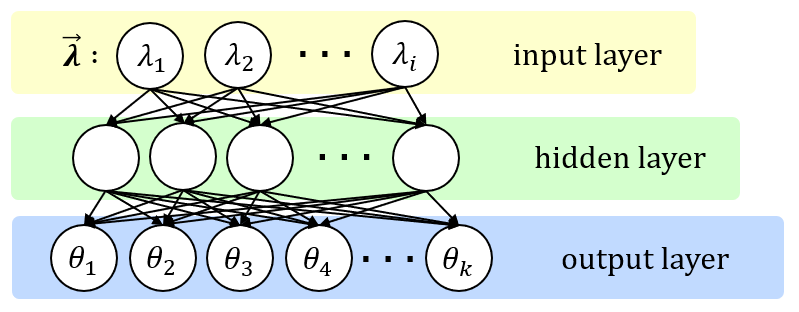}{Neural network of the NN-VQE. The neural network has one input layer, one hidden layer, and one output layer, which are all fully-connected linear layers. The number of nodes of the input layer corresponds to the number of Hamiltonian parameters. The node number of the hidden layer is a hyperparameter that varies with the circuit depth $D$. The node number of the output layer corresponds to the number of PQC parameters.}{fig:supp_nn}{width=0.55\linewidth}
        
        \subsection{Technical details for the simulation}

            For Fig. 2, the Hamiltonian is an 8-qubit one-parameter 1D XXZ model. The training set of $\Delta$ is composed of 20 equispaced points in the interval of $[-3.0, 3.0]$; the test set is 201 equispaced values of $\Delta$ in the interval of $[-4.0, 4.0]$. The circuit ansatz is 8-qubit MERA with depth $D = 1, 2, 3$. When $D = 1$, the node number of the hidden layer for the neural network is $20$, and the node number of the output layer corresponds to the number of PQC parameters $74$, without dropout and with $dropout = 0.30$; when $D = 2$, the node number of the hidden layer is $20$, and the node number of the output layer corresponds to the number of PQC parameters $124$, without dropout and with $dropout = 0.05$; when $D = 3$, the node number of the hidden layer is $30$, and the node number of the output layer corresponds to the number of PQC parameters $174$, without dropout and with $dropout = 0.20$. The starting learning rate for the optimizer is $0.009$, and it decays to its $70\%$ every $1000$ steps. The maximum iteration for the optimization is $2500$.
            
            When introducing the active learning strategy as shown in Fig. 3, the Hamiltonian we used is an 8-qubit one-parameter 1D XXZ model. The training set of $\Delta$ is 11 actively learned points in the interval of $[-3.0, 3.0]$; the test set is 201 equispaced values of $\Delta$ in the interval of $[-3.0, 3.0]$. The ansatz is 8-qubit MERA with $D = 2$. The node number of the hidden layer is $25$, and the node number of the output layer corresponds to the number of PQC parameters $124$, without dropout and with $dropout = 0.20$. The hyperparameter $\mu$ in the cost function of active learning is 6.0. The starting learning rate is $0.009$, and it decays to its $85\%$ every $200$ steps. The maximum iteration is $2500$.
            
            When showing the speedup in the training process as shown in Fig. 4, the Hamiltonian is a 12-qubit one-parameter 1D XXZ model. The training set is $\Delta = 1.5$ or $\Delta = 2.0$ with only one point. The circuit ansatz is 12-qubit HEA with $D = 3$. The node number of the hidden layer is $36$, and the node number of the output layer corresponds to the number of PQC parameters $216$, with $dropout = 0.20$. The learning rate is $0.009$. The maximum iteration is $100$.
            
            For the result in Fig. 5, the Hamiltonian is a 12-qubit two-parameter 1D XXZ model. The training set of $\Delta$ is composed of 10 equispaced points in the interval of $[-1.0, 1.0]$, and $\lambda$ is composed of 5 equispaced points in the interval of $[0.0, 1.0]$; the test set of $\Delta$ is composed of 101 equispaced points in the interval of $[-1.0, 1.0]$, and $\lambda$ is composed of 51 equispaced points in the interval of $[0.0, 1.0]$, The ansatz is 12-qubit HEA with $D = 1, 2$; the node number of the hidden layer is chosen at $40$, and the node number of the output layer corresponds to the number of PQC parameters ($96$ when $D = 1$ and $156$ when $D = 2$), with $dropout = 0.2$. The starting learning rate is $0.01$, and it decays to its $70\%$ every $800$ steps. The maximum iteration is $4000$.

        \subsection{Result of two-parameter 1D XXZ model when $n = 8$}
        
\ipic{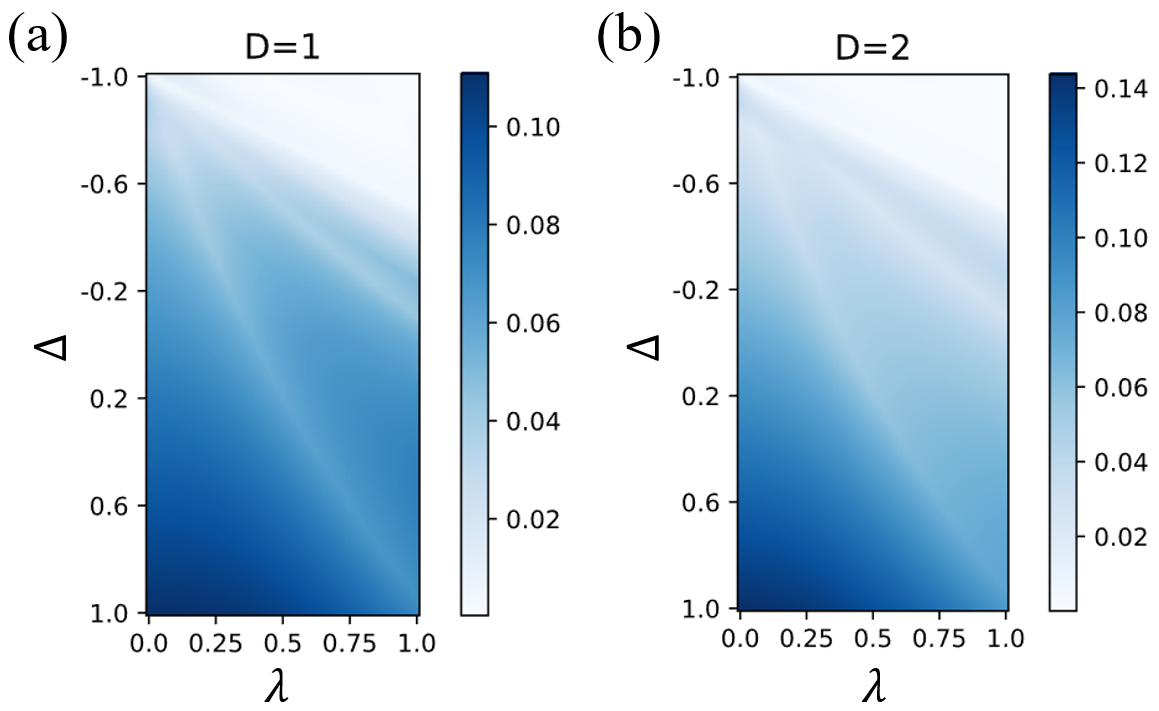}{Relative errors of ground-state energies for an $n = 8$ 1D XXZ spin chain with two tunable Hamiltonian parameters, using hardware-efficient ansatz with circuit depth $D = 1, 2$.}{fig:supp_2xxz}{width=0.6\linewidth}

            In this section, we use a two-parameter 8-qubit 2D XXZ model as the Hamiltonian. The training set of $\Delta$ is composed of 10 equispaced points in the interval of $[-1.0, 1.0]$, and $\lambda$ is composed of 5 equispaced points in the interval of $[0.0, 1.0]$; the test set of $\Delta$ is composed of 101 equispaced points in the interval of $[-1.0, 1.0]$, and $\lambda$ is composed of 51 equispaced points in the interval of $[0.0, 1.0]$; the ansatz is 8-qubit HEA with $D = 1, 2$; the node number of the hidden layer is chosen at $25$, and the node number of the output layer corresponds to the number of PQC parameters ($64$ when $D = 1$ and $104$ when $D = 2$), $dropout = 0.2$. The starting learning rate is $0.01$, and it decays to its $70\%$ every $800$ steps. The maximum iteration is $4000$.

            The numerical results are presented in Fig. \ref{fig:supp_2xxz}. Same as $n = 12$ result in the main text, the NN-VQE with a two-input neural network shows excellent performance in estimating ground-state energy, which highlights the effectiveness of our approach in estimating multiple-parameter Hamiltonian.

        \subsection{A comparison between NN-VQE and meta-VQE}

\ipic{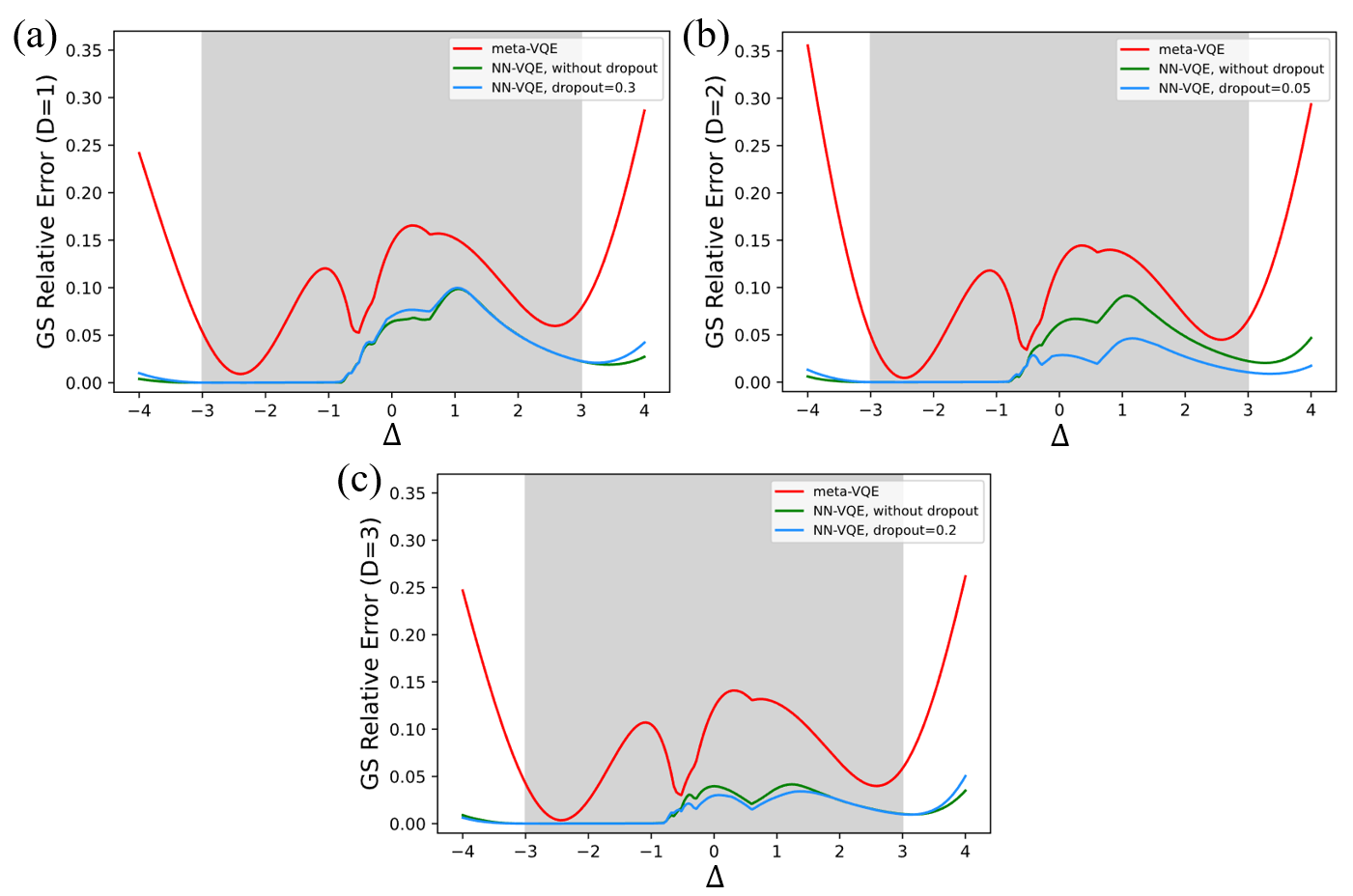}{A comparison between NN-VQE and meta-VQE. The parameterized Hamiltonian is an 8-qubit one-parameter 1D XXZ spin chain with a transverse field strength fixed at $\lambda = 0.75$. The ansatz is MERA with (a) $D = 1$, (b) $D = 2$, and (c) $D = 3$. The training set interval is on the gray background and the whole line is in the interval of the test set.}{fig:supp_comparison}{width=0.83\linewidth}

\ipic{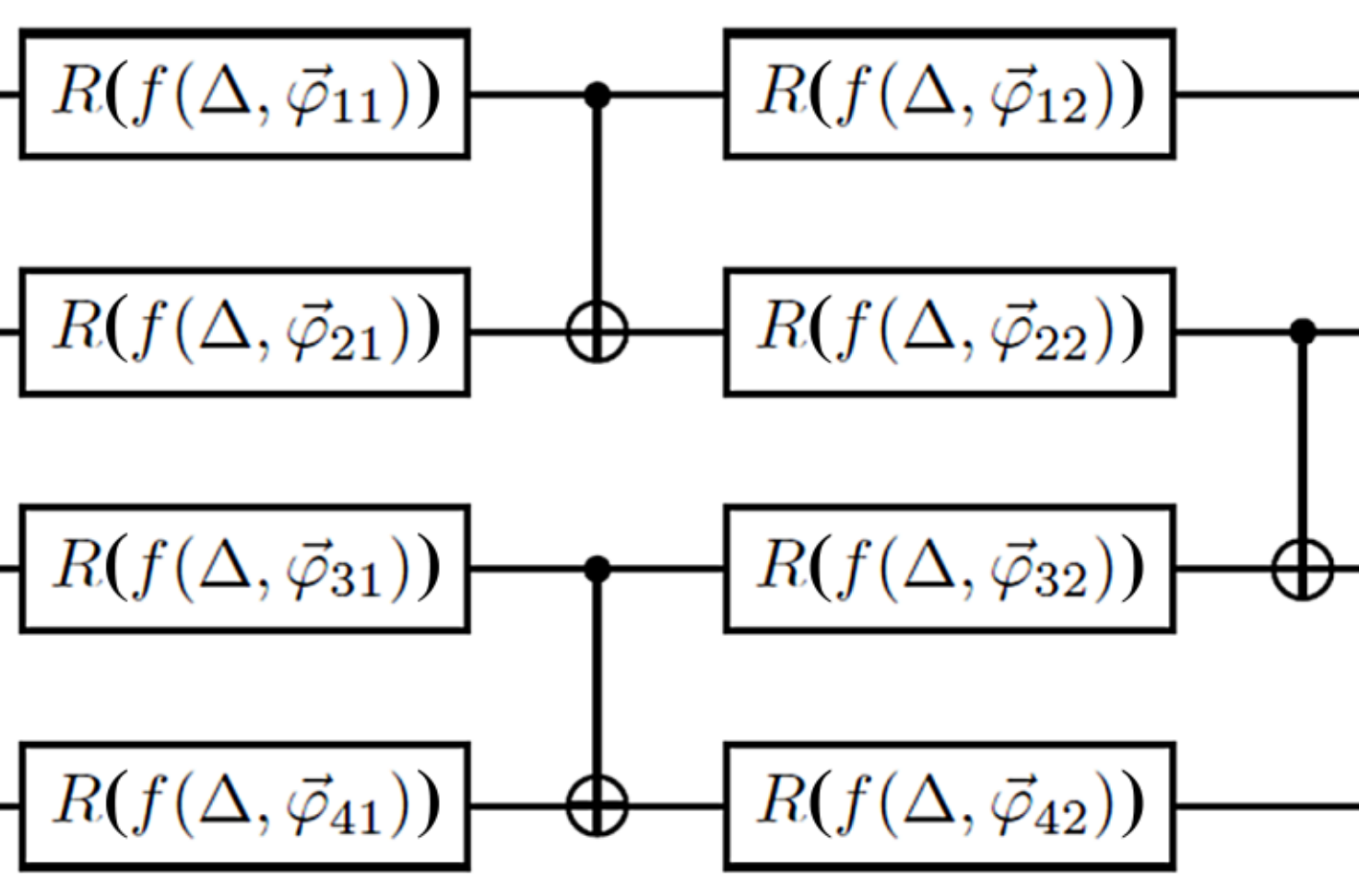}{The 4-qubit example encoding layer of meta-VQE from \cite{Cervera-Lierta2021}. Each $R(\bm{\theta})$ gate corresponds to $R_z(\theta_{1})$$R_x(\theta_{2})$. The function used for the encoding layer is $f(\Delta, \phi) = w\Delta + \phi$.}{fig:supp_meta_encoding}{width=0.35\linewidth}

            We use the 8-qubit one-parameter 1D XXZ model as the Hamiltonian which is the same Hamiltonian as in the meta-VQE work. The training set of $\Delta$ is composed of 20 equispaced points in the interval of $[-3.0, 3.0]$, and the test set is 201 equispaced values of $\Delta$ in the interval of $[-4.0, 4.0]$.
            
            The NN-VQE circuit ansatz and the meta-VQE processing layer are set the same as the MERA circuit. The encoding layer of meta-VQE is the same as \cite{Cervera-Lierta2021} (see Fig. \ref{fig:supp_meta_encoding}). The encoding function it uses is $f(\Delta, \phi) = w\Delta + \phi$, where $\Delta$ is the Hamiltonian parameter and $w$ and $\phi$ are encoding parameters to optimize.

            As for the neural network of the NN-VQE (see Fig. \ref{fig:supp_nn}), we choose the hidden layer node number $=20$ when $D = 1, 2$, and the hidden layer node number $=30$ when $D = 3$. Dropout was also employed to avoid overfitting. 
            The starting learning rate is $0.009$, and it decays to its $70\%$ every $1000$ steps. The max iteration is $2500$.

            The result of meta-VQE and NN-VQE (with and without dropout) is shown in Fig. \ref{fig:supp_comparison} with different $D$s. In the region of the training set, NN-VQE performs far better than meta-VQE, especially when the circuit gets deeper. When it comes to the training points not in the test region, the error of our NN-VQE is significantly lower than that of the meta-VQE. The result shows that our NN-VQE can give more precise ground-state energy estimation than meta-VQE and has better generalization ability without any fine-tuning.

\ipic{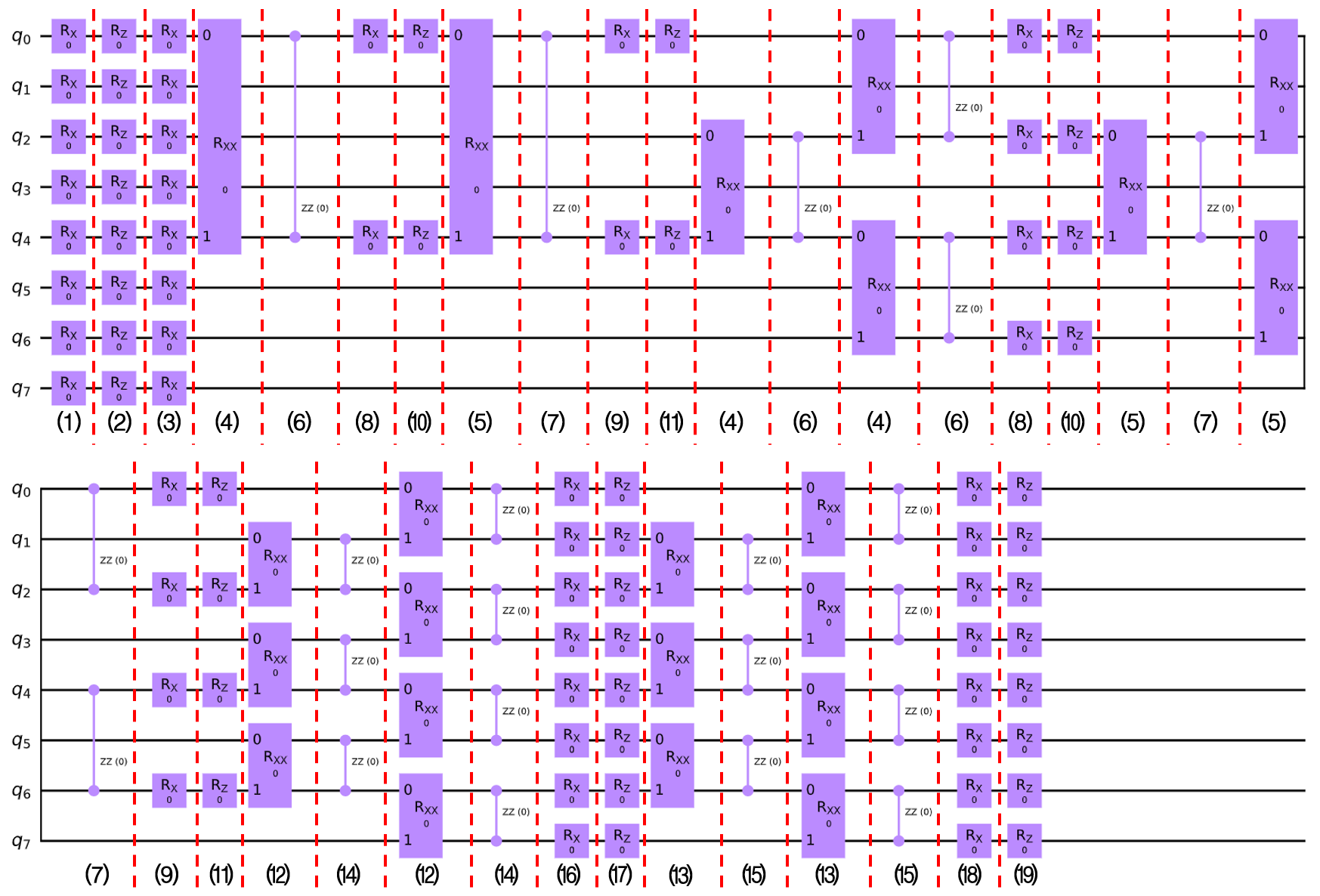}{$n = 8$ MERA circuit with $D = 2$.}{fig:supp_mera_2}{width=0.9\linewidth}

\ipic{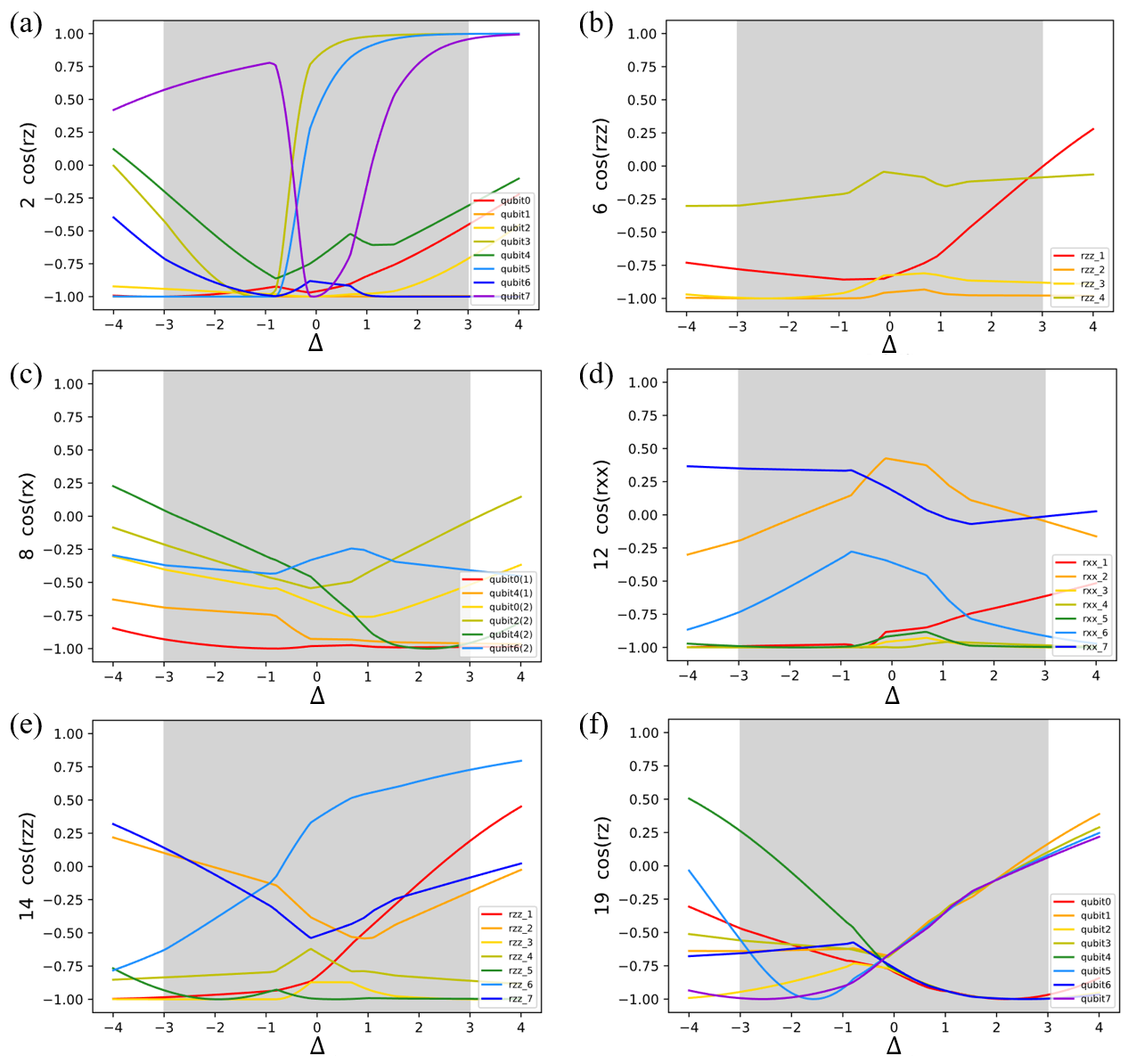}{Parameters of gates with respect to $\Delta$. The ansatz used is MERA with $D=2$. The number of each subplot corresponds to a set of gate numbers in Fig. \ref{fig:supp_mera_2}. Qubit numbers in the labels of the subplots associated with the qubits (have gates) in order, arranged from left to right and from top to bottom.}{fig:supp_gates_params}{width=0.85\linewidth}

        \subsection{Circuit parameters change visualization}
        
            To gain deeper insights, we examine the variations of gate parameters by plotting their cosine values with respect to the Hamiltonian parameter $\Delta$, shown in Fig. \ref{fig:supp_gates_params}. The sequence number of the subfigures corresponds to the gate number in Fig. \ref{fig:supp_mera_2}. As shown in the figures, the relationship between $\Delta$ and the gate parameters (or the cosine of the gates' parameters) is neither linear nor low-order polynomial, which could explain why our NN-VQE performs far better than meta-VQE that utilizes simple analytical formula \cite{Cervera-Lierta2021}. The abrupt change of the circuit parameters is often related to the region near criticality (see Fig. \ref{fig:supp_phase}). Instead, smooth variations are observed on the ferromagnetic (FM) phase and the antiferromagnetic (AFM) phase. Consequently, this non-trivial relationship indicates that employing a neural network as the encoding module is of great necessity to capture various quantum phases with the same setup.

            The model investigated is the MERA circuit with $D = 2$ and $dropout = 0.05$. 

        \subsection{Additional results on the optimization process speed-up by NN-VQE}
        
\begin{table}[!h]
    \centering 
    \begin{tabular}{|c|r|r|r|r|r|r|r|r|r|} \hline 
        $\Delta$ & \multicolumn{3}{|c|}{1.0} & \multicolumn{3}{|c|}{1.5} & \multicolumn{3}{|c|}{2.0} \\ \hline
        $n$ & \multicolumn{1}{|c|}{8} & 10 & 12 & \multicolumn{1}{|c|}{8} & 10 & 12 & \multicolumn{1}{|c|}{8} & 10 & 12 \\ \hline
        NN-VQE / \% & 60 & 60 & 70 & 80 & 70 & 60 & 90 & 85 & 85 \\ \hline
        VQE / \% & 5 & 0 & 20 & 5 & 5 & 0 & 20 & 30 & 10 \\ \hline
    \end{tabular}
    \caption{Convergent rate. The convergent threshold of ground-state energy relative error is set to $0.1$, we call the model converged when the energy error is below such a value and within the optimization epoch budget. We call the number of convergent models divided by total trials as the convergent rate. The convergence rates of an XXZ spin chain with $\Delta = 1.0, 1.5, 2.0$ and $n = 8, 10, 12$ within the 100 epochs are reported. The PQC ansatz was the hardware-efficient ansatz with $D = 3$, and the optimizer is Adam. }
    \label{table:supp_con_rate}
\end{table}

\ipic{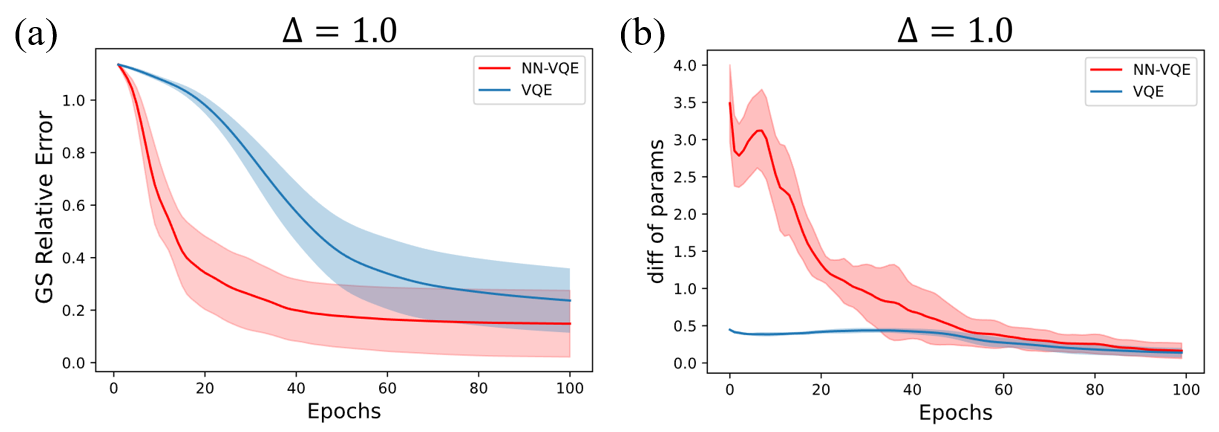}{Speedup in the optimization process of NN-VQE and the corresponding PQC parameter changes. (a) The ground-state energy relative errors for an $n = 12$ XXZ spin chain when $\Delta = 1.0$ are shown with respect to epochs. Hardware-efficient ansatz with $D = 3$ is used. The red and blue lines correspond to the ground-state energy relative errors and standard deviation of NN-VQE and VQE respectively. (b) The summed parameter updates when training VQE and NN-VQE in corresponding $\Delta$. }{fig:speedup_1.0}{width=0.75\linewidth}

        
            It is mentioned that NN-VQE could speed up the training process, allowing rapid convergence. Here we use the same ansatz circuit, Hamiltonian, and training schedule as in the main text but report further details.

            To avoid the impact of parameter initialization, the initial PQC parameters of the standard VQE are set the same as the initial PQC parameters of the NN-VQE, which are generated by a randomly initialized neural network. The results of $\Delta = 1.5, 2.0$ are shown in the main text and we show the supplemental result for $\Delta = 1.0$ in Fig. \ref{fig:speedup_1.0}. We can see that NN-VQE converges much faster than standard VQE when training. Also, NN-PQC hybrid architecture brings a more dramatic change in the PQC parameters at the beginning stage of the optimization process which might be helpful in escaping barren plateaus.

            Furthermore, We investigate the results of different $\Delta$s and  system sizes $n$s. The result is summarized in Table \ref{table:supp_con_rate}. We can see that the convergent rate of the NN-VQE is higher than VQE the number of epoch budgets is 100.

            The setup hyperparameters for Tab. S1 is as follows. When $n = 8$, the ansatz is 8-qubit MERA with $D = 3$. The node number of the hidden layer is $25$, and the node number of the output layer corresponds to the number of PQC parameters $144$, with $dropout = 0.20$. The learning rate is $0.009$. The max iteration is $100$.
            When $n = 10$, the ansatz is 8-qubit MERA with $D = 3$. The node number of the hidden layer is $32$, and the node number of the output layer corresponds to the number of PQC parameters $180$, with $dropout = 0.20$. The learning rate is $0.009$. The max iteration is $100$.

        \subsection{Quantum software framework}
        
            All the high-performance numerical simulations in this work are conducted with TensorCircuit \cite{Zhang2023}: an open-source, high-performance, full-featured quantum software framework for the NISQ era. The software simulates the quantum circuit with an advanced tensor network contraction engine and supports modern machine learning engineering paradigms: automatic differentiation, vectorized parallelism, just-in-time compilation, and GPU acceleration. It is specifically suitable to simulate the hybrid system with both neural networks and quantum circuits.

        \subsection{The phase diagram of the 1D XXZ model}
        
\ipic{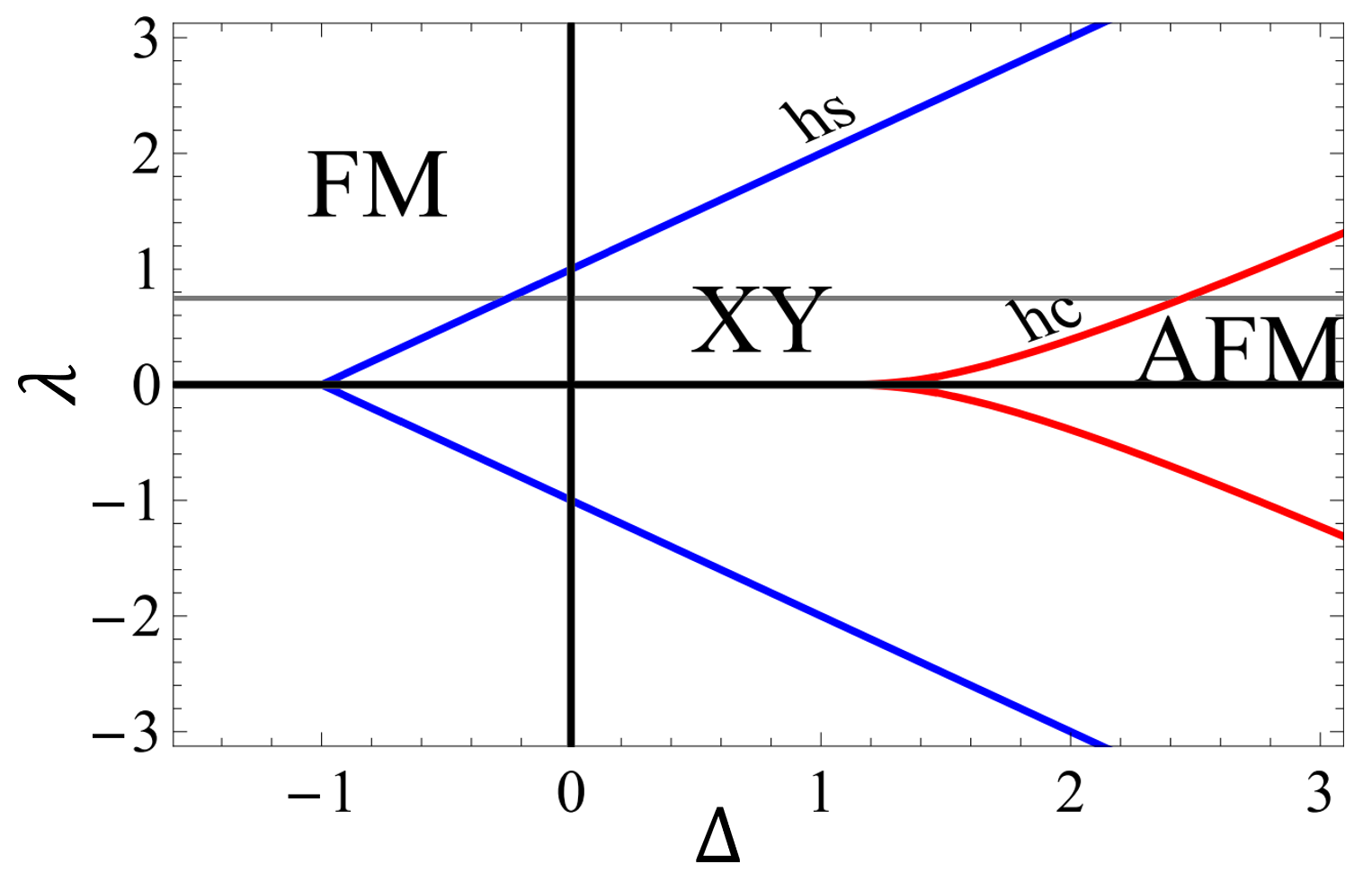}{Phase diagram of the 1D XXZ model \cite{braiorrorrs2015}. The line $hs$ separates the XY phase from the ferromagnetic (FM) phase. The curve $hc$ separates the anti-ferromagnetic (AFM) phase from the XY phase. The gray line $\lambda = 0.75$.}{fig:supp_phase}{width=0.45\linewidth}

            The phase diagram of the 1D XXZ model is shown in Fig. \ref{fig:supp_phase}. The three phases are separated by two curves $hs$ and $hc$
            \begin{eqnarray}
                hs &=& d(1+\Delta) \\
                hc &=& \frac{\pi \sinh \lambda}{\lambda} \sum_{n=-\infty}^{\infty} {\rm sech} \frac{\pi^2}{2\lambda} (1+2n),
            \end{eqnarray}
            where $d = 1$ is the dimension of the model and $\lambda = {\rm arccosh} \Delta$.

        \subsection{Related works}

            Meta-VQE \cite{Cervera-Lierta2021} aims to solve the problem of parameterized Hamiltonian. They divide the PQC into two parts: the encoding layer and the processing layer. In the encoding layer, Hamiltonian parameters are mapped to some simple formula. And the processing layer is a standard VQE requiring fine tuning for each Hamiltonian parameter. The method has some drawbacks though. Firstly, most of the circuit parameters still need to be trained or fine-tuned on separate Hamiltonian instances, and secondly, the expressive power of their parameter prediction model is very weak to achieve the given accuracy. On the contrary, our NN-VQE encodes all parameters in the circuit and utilizes a more complicated neural network for the prediction. These improvements greatly improve the performance and accuracy of the method due to the high expressiveness of neural networks. Besides, our method totally avoids retraining or fine-tuning on each separate Hamiltonian instance (See Fig. \ref{fig:supp_comparison}) since all circuit parameters are encoded. 

            In this work, we implement VQE as a representative example of VQAs to demonstrate the efficacy of our NN-VQA framework, but other NN-VQAs for different applications can also show their effectiveness. Some works employed similar ideas in this work for optimization algorithm (QAOA). Since identifying the optimal parameters is a difficult task in QAOA, researchers employ neural networks or other machine learning techniques to find better initialization parameters. A spectrum of neural network architectures, including Deep Neural Networks (DNN) \cite{amosy2022}, Convolutional Neural Networks (CNN) \cite{xie2023}, and Graph Neural Networks (GNN) \cite{Jain2022} have been explored. By employing these neural network encoding methods, significant progress has been made in enhancing the performance and efficiency of QAOA. However, in the QAOA cases, the neural network encoded parameters are only utilized as a good initialization point and further fine-tuning on QAOA is still required. On the contrary, in the NN-VQE case, we directly use the neural network encoded parameters as the final parameter choice which has already given satisfying performance in real applications. 

            Our work also shares some similarities with \cite{Wang2022}. In their work, with the Hamiltonian parameters as input to the GNN, they extract some intermediate representation as the input for another generative network to generate the classical shadows \cite{Huang2020b} of the ground state, which is an efficient but approximate representation of the quantum state. In our framework, we also extract the intermediate representation via encoding neural network and Hamiltonian as input. However, we aim to directly generate the ground state in the quantum form on a quantum computer instead of only recovering a classical shadow representation with many known limitations.

\end{widetext}
\end{document}

%% file: main.bbl
%